\documentclass[12pt]{article}             % Use with LaTeX2e
\usepackage{graphicx} % For png figures [pdftex]
\usepackage{wrapfig}
\begin{document}

\bibliographystyle{unsrt}
\begin{center}
{\Large\bf  Canonical transforms, quantumness and probability
representation of quantum mechanics}\\\vspace{2mm}
{\bf Margarita A. Man'ko and Vladimir I. Man'ko}\\
{\it P. N. Lebedev Physical Institute\\
Leninskii Prospect 53, Moscow 119991, Russia}\\
e-mails: ~~~mmanko@sci.lebedev.ru,~~~manko@sci.lebedev.ru
\end{center}

\vspace{2mm}

\begin{abstract}
The linear canonical transforms of position and momentum are used to
construct the tomographic probability representation of quantum
states where the fair probability distribution determines the
quantum state instead of the wave function or density matrix. The
example of Moshinsky shutter problem is considered.
\end{abstract}

 \vspace{0.4cm}

\section*{Introduction}
Both in classical and quantum mechanics, the linear canonical
transforms of position and momentum preserving the Poisson brackets
and commutation relations, respectively, were studied in
\cite{MoshQuesne,SeligmanKramer}. The linear canonical transform of
creation and annihilation operators in the form of Bogolubov
transform~\cite{Bogolubov} can be associated with squeezing
phenomenon~\cite{Pleban} and correlated squeezed
states~\cite{KurmishevPhysLett1980}. Canonical transforms are also
important for considering linear integrals of motion for quadratic
nonstationary systems~\cite{Malkin,Dod183,Sel-Franc} and light
propagation within the framework of geometric
optics~\cite{Wolfbook,WolfLiemethodsinOptics}.

Recently \cite{ManciniPLA,IbortPhysScr} the new formulation of
quantum mechanics (the probability representation of quantum
mechanics) was suggested, where the quantum states are described by
fair probability distributions called symplectic tomograms. The
tomograms are related to the Wigner function~\cite{Wig32} of quantum
states by means of integral Radon transform~\cite{Radon}. In
\cite{MoshinskyPRA}, the problem of diffraction in
time~\cite{Mosh1946} connected with quantum shutter was solved
within the framework of the probability representation of quantum
mechanics.

The aim of this work is to point out the connection of linear
canonical transform of position and momentum with quantum tomograms
describing the quantum  states and to discuss the quantumness and
classicality of the system states.

The paper is organized as follows.

In Section 1, we review the tomographic probability representation.
In Section~2, we consider the problem of Moshinsky shutter in the
phase-space representation. In Section~3, the conclusions and
prospectives are presented.

\section{Symplectic tomography}
The symplectic tomogram can be constructed using the star-product
scheme suggested in \cite{OVMARMO1,OVMARMO4} with a pair of
operators, which are the quantizer operator
\begin{equation}\label{eq.1}
\hat{\cal
D}(X,\mu,\nu)=\frac{1}{2\pi} \exp\left(iX\hat1-i\nu\hat p-i\mu\hat
q\right)
\end{equation}
and the dequantizer operator
\begin{equation}\label{eq.2}
 \hat{\cal U}(X,\mu,\nu)=\delta(X\hat 1-\mu\hat
q-\nu\hat p), \end{equation} where $X$, $\mu$, and $\nu$ are real
variables, and $\hat q$ and $\hat p$ are the position and momentum
operators, respectively.

The symplectic tomogram of quantum state with the density operator
$\hat \rho$ is defined as
\begin{equation}\label{eq.3}
w\left (X,\,\mu ,\,\nu \right )=\langle\hat{\cal
U}(X,\mu,\nu)\rangle=\mbox{Tr}\left(\hat\rho \delta (X -\mu \hat q
-\nu \hat p)\right ), \qquad \hbar=1.
\end{equation}
The tomogram can be rewritten in terms of Radon transform of the
Wigner function $W(q,p)$ as follows:
\begin{equation}\label{eq.4}
 w\left(X,\,\mu,\,\nu\right )=\frac{1}{2\pi}\int
W(q\,,p)\delta\,(X-\mu q-\nu p)~dq~dp.
\end{equation}
The inverse Radon transform reads
\begin{eqnarray}\label{eq.5}
W(q,\,p)=\frac {1}{2\pi }\int w\left (X,\,\mu ,\,\nu \right )\exp
\left [-i\left (\mu q+\nu p-X\right )\right ] \,d\mu \,d\nu \,dX.
\end{eqnarray}
The tomogram is normalized, i.e.,
\[\int w\left (X,\,\mu ,\,\nu \right)d X=1\]
for all $\mu$ and $\nu$. Relation (\ref{eq.4}) corresponds to
operator form of the inverse Radon transform  for the density
operator
\begin{equation}\label{eq.6}
\hat\rho=\frac{1}{2\pi}\int w(X,\mu,\nu)e^{i(X-\mu\hat q-\nu\hat
p)}\,dX\, d\mu\, d\nu.
\end{equation}
The symplectic tomogram is the probability distribution of random
position $X$ measured in a specific reference frame in the phase
space. The reference frame is determined by two real parameters
$\mu=s\cos\theta$ and $\nu=s^{-1}\sin\theta$, where $s$ is the
scaling parameter and $\theta$ is the rotation angle of the frame
axes.

The tomographic symbol $f_A(X,\mu,\nu)$ of any operator $\hat A$ can
be given as $\mbox{Tr}\,\hat A \hat{\cal U}=f_a(X,\mu,\nu)$.

We consider the operator $|\psi_1\rangle\langle\psi_2|$, where
$|\psi_1\rangle=\exp(-i\hat H t_1)|\psi\rangle$ and
$|\psi_2\rangle=\exp(-i\hat H t_2)|\psi\rangle$. Here the
Hamiltonian is the sum of the kinetic and potential energies: $\hat
H=({\hat p^2}/{2})+U(\hat q).$ We use dimensionless units,
$\hbar=m=1$. The operator $\hat A=|\psi_1\rangle\langle\psi_2|$ for
$t_1=t_2=t$ is the density operator
\begin{equation}\label{eq.7}
\hat\rho(t)=\exp(-i\hat H t)|\psi\rangle\langle\psi|\exp(i\hat H t)
\end{equation}
of the system state.

The Schr\"odinger equation in the position representation for the
wave function $\psi(x_1,t_1)=\langle x_1|\psi_1\rangle$ provides the
equation for the function $\langle x_1|\hat A|
x_2\rangle=A(x_1,x_2,t_1,t_2)=\psi(x_1,t_1)\psi^\ast(x_2,t_2)$ in
the form
\begin{equation}\label{eq.8}
\left[i\left(\frac{\partial}{\partial t_1}+\frac{\partial}{\partial
t_2}\right)+\frac{1}{2}\left(\frac{\partial^2}{\partial
x_1^2}-\frac{\partial^2}{\partial
x_2^2}\right)+U(x_2)-U(x_1)\right]A(x_1,x_2,t_1,t_2)=0.
\end{equation}
At $t_1=t_2=t$, this equation is the standard von Neumann equation
for the density matrix in the position representation. This equation
can be rewritten for Weyl symbol of the operator $\hat A$,
\begin{equation}\label{eq.9}
W_A(q,p,t_1,t_2)=\mbox{Tr}\left(2\exp[2(\alpha\hat
a^+-\alpha^\ast\hat a)]\hat P\hat A\right),
\end{equation}
where $\alpha=({q+ip})/{\sqrt2}$, $\hat a=({\hat q+i\hat p})/{\sqrt
2}$, and $\hat P$ is the parity operator.

For $t_1=t_2=t$, the symbol $W_A(q,p,t_1=t,t_2=t)$ coincides with
the Wigner function $W(q,p,t)$ of the quantum state. The equation
reads
\begin{equation}\label{eq.10}
\left[i\left(\frac{\partial}{\partial t_1}+\frac{\partial}{\partial
t_2}\right)+i p\frac{\partial}{\partial
q}-U\left(q+\frac{i}{2}\frac{\partial}{\partial
p}\right)+U\left(q-\frac{i}{2}\frac{\partial}{\partial
p}\right)\right]W_A(q,p,t_1,t_2)=0.
\end{equation}
This equation provides the Moyal equation \cite{Moyal} for the
Wigner function.

Equation (\ref{eq.8}) can be also rewritten in the tomographic form
using the symbol of operator $\hat A$,
\begin{equation}\label{eq.11}
w_A(X,\mu,\nu,t_1,t_2)=\mbox{Tr}\left(\delta(X-\mu\hat q -\nu\hat
p)|\psi_1\rangle\langle\psi_2|\right).
\end{equation}
This function satisfies the equation
\begin{equation}\label{eq.12}
\left[i\left(\frac{\partial}{\partial t_1}+\frac{\partial}{\partial
t_2}\right)
-i\mu\frac{\partial}{\partial\nu}-\left(U\left(-\left(\frac{\partial}{\partial
X}\right)^{-1}\frac{\partial}{\partial\mu}+\frac{i}{2}\nu\frac{\partial}{\partial
X}\right) -\mbox{c.c.}\right)\right]w(X,\mu,\nu,t_1,t_2)=0.
\end{equation}
For $t_1=t_2$, the symbol of operator $\hat A$ becomes the
quantum-state tomogram $w(X,\mu,\nu,t)$ which satisfies the
evolution equation found in \cite{ManciniPLA}.

For harmonic oscillator, Eq.~(\ref{eq.12}) reads
\begin{equation}\label{eq.13}
\left[\left(\frac{\partial}{\partial t_1}+\frac{\partial}{\partial
t_2}\right)
-\mu\frac{\partial}{\partial\nu}+\nu\frac{\partial}{\partial\mu}
\right]w(X,\mu,\nu,t_1,t_2)=0.
\end{equation}

The quantumness and classicality of the system states can be
formulated in terms of the tomograms as follows.

A given normalized nonnegative tomogram $w(X,\mu,\nu)$ satisfying
the homogeneity condition $~w(\lambda X,\lambda \mu,\lambda
\nu)=|\lambda|^{-1}w(X,\mu,\nu)$, which follows from the homogeneity
property of the delta-function, corresponds to a quantum state iff
the integral on the right-hand side of Eq.~(\ref{eq.6}) is
nonnegative operator. The tomogram satisfying the nonnegativity of
the integral term in Eq.~(\ref{eq.5}) corresponds to the classical
state. The tomograms violating nonnegativity conditions of integrals
in both Eqs.~(\ref{eq.5}) and (\ref{eq.6}) correspond neither
classical nor quantum states.

The quantumness condition can be expressed also in the form of
entropic inequality \cite{M.M.Found.Phys.}
\begin{eqnarray}\label{eq.14}
&&-\left[\int w(X,\cos\theta,\sin\theta)\ln w(X,\cos\theta,\sin\theta) d X\right.\nonumber\\
&&\left.+\int w(X,\sin\theta,-\cos\theta)\ln
w(X,\sin\theta,-\cos\theta)d X\right]\geq\ln\pi e \label{eq.14}
\end{eqnarray}
for optical tomogram $w(X,\mu=\cos\theta,\nu=\sin\theta).$

Inequality (\ref{eq.14}) can be violated in the classical domain but
must be fulfilled in the quantum domain.

\section{Moshinsky shutter and diffraction in time}
In \cite{Mosh1946}, Moshinsky considered the problem of diffraction
in time. This problem corresponds to opening at time $t=0$
completely absorbing shutter located at position $x=0$, on which a
stream of particles of definite momentum $K$ was inpunged. The
Schr\"odinger equation (in dimensionless units)
\[i\dot\psi=-\frac{1}{2}\frac{\partial^2}{\partial x^2}\psi\]
was solved in terms of error function which for the shutter problem
is reduced to the Moshinsky function
\begin{equation}\label{M1}
M(x,k,t)=\frac{i}{2\pi}\int\frac{\exp[i(\kappa
x-[\kappa^2t/2])]}{\kappa-k}d\kappa.
\end{equation}
The expression $|M(x,k,t)|^2$ gives the probability density of
finding the particle at point $x$ at time $t$, if initially it was
on the left side of the shutter. Equation~(\ref{M1}) gives (see,
e.g., \cite{MoshinskyPRA})
\begin{equation}\label{M2}
|M(x,k,t)|^2=\frac{1}{2}\left\{\left[\frac{1}{2}-C(w)\right]^2+\left[\frac{1}{2}-S(w)\right]^2\right\}.
\end{equation}
Here $C(\omega)$ and $S(\omega)$ are Fresnel integrals
\begin{equation}\label{M3}
C(\omega)=\sqrt{\frac{2}{\pi}}\int_0^w\cos y^2d y,\qquad
S(\omega)=\sqrt{\frac{2}{\pi}}\int_0^w\sin y^2d y.
\end{equation}
One can find also the solution to the Moyal equation for the shutter
problem, and it provides the Wigner function of the
problem~\cite{MoshinskyPRA} in the form
\begin{equation}\label{M4}
W(q,p,k,t)=\frac{1}{\pi(k-p)}\sin\{2(p t-q)(k-p)\}\theta(p t-q).
\end{equation}

The symplectic tomogram for the Moshinsky shutter problem can also
be obtained as a solution to the tomographic evolution equation
(\ref{eq.12}); for $t_1=t_2=t$, it reads~\cite{MoshinskyPRA}
\begin{equation}\label{M5}
w(X,\mu,\nu,t)=\frac{1}{2|\mu|}\left\{\left[\frac{1}{2}+C(\rho)\right]^2+\left[\frac{1}{2}+
S(\rho)\right]^2\right\},
\end{equation}
where
\begin{equation}\label{M6}
\rho=\frac{k(\mu t+\nu)-X}{\sqrt{2\mu(\mu t+\nu)}}.
\end{equation}
Thus, the shutter problem can be solved in the Schr\"odinger, Moyal,
and tomographic-probability representations. One can check that the
solutions of these equations contain information on the
diffraction-in-time properties of the shutter. The solutions are
related by the integral transforms. For example, tomogram~(\ref{M5})
and Wigner function~(\ref{M4}) satisfy (\ref{eq.5}).

\section{Canonical transforms and tomography}
In Eq.~(\ref{eq.4}), the argument of delta-function provides the
classical linear transform of the particle position
$q\rightarrow\mu q+\nu p$. This transform together with the linear
transform of the particle momentum $p\rightarrow\mu' q+\nu' p$
form canonical transform in the phase space preserving the Poisson
brackets. The real 2$\times$2 matrices
\[
\Lambda=\left(\begin{array}{cc}
\mu&\nu\\
\mu'&\nu'\end{array} \right),
\]
with the determinant equal to unity, form the symplectic group
Sp(2,R).

In the quantum phase space, there exists the transform of operators
$\hat q\rightarrow\mu\hat q+\nu \hat p$, $\hat p\rightarrow\mu'\hat
q+\nu'\hat p$ determined by the symplectic matrix $\Lambda$. This
transform corresponds to the unitary irreducible representation of
the symplectic group. It can be considered as a combination of the
scaling transform $~\hat S=\exp i s\left(\hat q\hat p+\hat p\hat
q\right)/2$ and rotation $~\hat R=\exp i\theta\left(\hat p^2+\hat
q^2\right)/2$.

The representations of the classical linear canonical transforms by
means of the quantum operators acting in the Hilbert space of the
particle states and the kernels of such operators were discussed in
\cite{MoshQuesne}. One can see that the quantizer and dequantizer of
the symplectic tomography star-product scheme are based on using the
quantum observables which are associated with the representations of
the classical canonical transforms, i.e., with the representations
of the symplectic group. An analogous construction can be given for
systems with many degrees of freedom.

\section*{Conclusions}
We point out the main results of our study.

We reviewed the notion of quantum state in the symplectic
tomographic probability representation.

We constructed the evolution equation for the tomographic symbol
of the operator corresponding to the product of the wave function
$\psi(x_1,t_1)$ and its complex conjugate $\psi^*(x_2,t_2)$.

We considered the problem of Moshinsky shutter in the probability
representation of quantum mechanics.

We formulated the quantumness of the system state as the operator
inequality and as the inequality for tomographic entropy.

An extension of this approach to multimode states will be presented
in future publications.

\subsection*{Acknowledgments}
The study was supported by the Russian Foundation for Basic Research
under Projects Nos.~09-02-00142 and 10-02-00312. The authors thank
the Organizers of the Conference in memory of Marcos Moshinsky
(Cuernavaca, Mexico, 2010) and especially Profs. Thomas Seligman and
Octavio Casta\~nos for invitation and kind hospitality.

\end{document}